# Phase and dispersion engineering of metalenses: broadband achromatic focusing and imaging in the visible

Wei Ting Chen[1], Alexander Y. Zhu[1], Vyshakh Sanjeev[1,3], Mohammadreza Khorasaninejad[1], Zhujun Shi[2], Eric Lee[1,3] and Federico Capasso[1,*]

[1]*Harvard John A. Paulson School of Engineering and Applied Sciences, Harvard University, Cambridge, Massachusetts 02138, USA*

[2]*Department of Physics, Harvard University, Cambridge, Massachusetts 02138, USA*

[3]*University of Waterloo, Waterloo, ON N2L 3G1, Canada*

[*]*Corresponding author:* capasso@seas.harvard.edu

**Metasurfaces have the potential to miniaturize and improve the performance of any optical element, with applications spanning telecommunications, computing and wearable optics. However, the ability to retain functionality over a continuous, broad range of wavelengths is essential for their widespread adoption. So far, efforts to achieve this have been limited to either a reflection configuration or operation at wavelengths other than the visible. Here, we show that by judicious choice of nanostructures, one can independently control both the phase and group delay, leading to broadband transmissive achromatic metalenses and metalenses with tunable dispersion across the visible spectrum. We demonstrate diffraction-limited**

**broadband focusing and achromatic imaging using metalenses across wavelengths from 470 nm to 670 nm, with focal length change of a few percent across this 200 nm bandwidth. This approach requires only a single layer metasurface whose thickness is on the wavelength scale, and does not involve spatial multiplexing or cascading.**

## Introduction

Conventional refractive optical components such as prisms and lenses are primarily manufactured by glass polishing techniques and impart their phases via volumetric propagation of light. In high-end, diffraction-limited optical systems the phase accuracy must be within a fraction of the incident wavelength. As a result, these traditional optical components are bulky, costly, and time-consuming to manufacture while maintaining high precision. This is a significant limitation, particularly for applications such as portable systems and wearable devices. In recent years, metasurfaces have emerged as a versatile platform for wavefront shaping. Since the phase is accurately controlled from 0 to $2\pi$ radians by subwavelength-spaced structures with thicknesses at the sub-wavelength or wavelength scale, many compact optical devices based on metasurfaces have been demonstrated. These include flat lenses [1-4], polarimeters [5-7], axicons [8, 9], polarization elements [10, 11] and holograms [12-15]. However, these devices are highly chromatic despite being usually comprised of weakly dispersive materials such as metals and dielectrics. This can be attributed to two separate factors: dispersion arising from periodic lattice (see Fig. S1 for detailed discussion), as well as dispersion arising from the individual structures due to their confinement of incident light in either a resonant or guided manner. Previous works have addressed this challenge by alternately using multiple coupled resonances to tailor phase profiles at several discrete frequencies [16, 17], stacking/stitching several layers of meta-surfaces [18], increasing the phase modulation to be more than 2π radians (giving rise to the so-called "multi-order diffractive lenses") [19] and engineering the dispersion [20-21]. Recently, several metalenses have been demonstrated capable of achromatic focusing with bandwidths of only a few tens of terahertz (THz) [22-25]. These works are mainly limited to near-infrared wavelengths, and experimental demonstration of achromatic imaging was not shown.

Here, we demonstrate achromatic metalenses covering nearly the entire visible range (from 470 nm to 670 nm, corresponding to a 200 THz bandwidth) with focal length change of the order of a few percent. Our approach is unique in that it relies on engineering group delay and group delay dispersion independently from the phase.

**Principle of achromatic metalenses**

As an initial example, consider an achromatic metalens shown in Fig. 1(a). For achromatic focusing at normal incidence, the relative phase $\varphi$ provided by the metalens elements with respect to the center follows [3]:

$$\varphi(r,\omega) = -\frac{\omega}{c}(\sqrt{r^2+F^2}-F) \qquad (1)$$

where $\omega$, $r$ and $F$ are angular frequency, radial coordinate, and focal length, respectively. The spatial- and frequency-dependent phase profile $\varphi(r,\omega)$ implies that at a given $r$, different transverse wavevectors $k_r = \frac{\partial \varphi(r,\omega)}{\partial r}$ must be provided by the metalens so that different wavelengths are deflected by the same angle. Equation 1 can be Taylor-expanded near a design frequency $\omega_d$ as

$$\varphi(r,\omega) = \varphi(r,\omega_d) + \left.\frac{\partial \varphi(r,\omega)}{\partial \omega}\right|_{\omega=\omega_d} (\omega - \omega_d) + \left.\frac{\partial^2 \varphi(r,\omega)}{\partial \omega^2}\right|_{\omega=\omega_d} (\omega - \omega_d)^2 + \ldots \quad (2)$$

Equation 2 indicates that to achieve achromatic focusing within a given bandwidth $\Delta\omega$ around $\omega_d$, an optical element placed at a radial coordinate *r* needs to satisfy not only the required relative phase ( $\varphi(r,\omega_d)$ ), but also the higher-order derivative terms. The derivative terms determine the metalens dispersion. $\frac{\partial \varphi(r,\omega)}{\partial \omega}$ and $\frac{\partial^2 \varphi(r,\omega)}{\partial \omega^2}$ are the relative group delay and group delay dispersion (GDD), and are typically of the order of femto-second (fs) and femto-second square ($fs^2$) in the visible. Conventional diffractive lenses only satisfy the required phase, i.e. the phase profile at a design frequency. Neglect of these derivative terms results in chromatic effects. An intuitive interpretation of each term of Eq. 2 is shown in Fig. 1(a). The first term leads to a spherical wavefront (yellow line in Fig. 1(a)). The group delay term compensates for the difference in the wavepackets' arrival times at the focus. The high-order derivative terms (GDD etc.) ensure that the wavepackets are identical. The net effect is minimization of the spread in the arrival times of wavepackets at the focus to ensure they constructively interfere. The smaller the time spread, the larger the bandwidth achievable. Therefore, in order to realize diffraction-limited focusing for a broad bandwidth, both phase and group delay, as well as higher order terms, need to be taken into account. One can see from Eq. (1) and (2) that, in an achromatic lens, the relative group delay satisfies:

$$\frac{\partial\varphi(r,\omega)}{\partial\omega} = -\frac{(\sqrt{r^2+F^2}-F)}{c} \qquad (3)$$

and the higher derivative terms are zero. The achievable range of group delay given by all elements is associated with the diameter and numerical aperture (NA) of the achromatic metalens.

To account for the dispersion of a metalens, the focal length *F* of Eq. 1 can be parametrized as:

$$F = k \times \omega^n \qquad (4)$$

where *k* is a positive constant and *n* is a real number. We refer to the metalenses with *n* = 0 and *n* = 1 as achromatic and diffractive metalenses hereafter. The diffractive metalens possesses a focal length that decreases with increasing wavelength, similar to Fresnel lenses. From Eq. 4, the positive and negative values of *n* imply that shorter/longer wavelengths are focused farther from/closer to the metalens, respectively. The larger the absolute value of *n*, the farther the separation between the focal spots of two wavelengths, resulting in stronger dispersion. Figure 1(b) and 1(c) show the required relative group delays and GDDs as a function of radial coordinate for metalenses with NA = 0.2 at $\lambda$ = 530 nm. Note that the NA is a function of wavelength for $n \neq 0$ due to the change in focal length. For $n$ = 2 and -1, they require non-negligible GDD to precisely control the focal length shift and achieve diffraction-limited focusing. Note that usually the required group delay and GDD are relatively small for $n$ = 1. This is in line with our previous observation that a diffractive metalens (NA = 0.8) implemented using geometric phase can still focus light with a focal spot size approximately equal to a wavelength [3]. In this paper, we consider dispersion terms up to GDD.

**Independent control of phase and group delay**

To expand the library of possible elements for the metalens and maximize the degrees of freedom in our design, we use two nano-fins for a single element and arrange them in a manner reminiscent of slot waveguides. The nano-fins' geometric parameters are defined in Fig. 2(a), and scanning electron microscopic images from a fabricated metalens are provided in Fig. 2(b) and Fig. S2. It has been previously shown that slot waveguides can support tunable dispersion, such as near zero GDD for a wide bandwidth [26, 27]. For simplicity and without loss of generality, we first consider the optical properties of a single $TiO_2$ nano-fin. The high aspect ratio $TiO_2$ nano-fin can be fabricated using electron beam lithography followed by atomic layer deposition [28]. When a left-handed circularly polarized beam (Jones vector notation $[1 \quad i]'$) passes through the nano-fin, the transmitted light can be described by [29]:

$$\frac{\tilde{t}_L + \tilde{t}_S}{2}\begin{bmatrix}1\\ i\end{bmatrix} + \frac{\tilde{t}_L - \tilde{t}_S}{2}\cdot \exp(i2\alpha)\begin{bmatrix}1\\ -i\end{bmatrix} \qquad (5)$$

where $\tilde{t}_L$ and $\tilde{t}_S$ represent complex transmission coefficients when the incident light is polarized along the long and short axis of the nano-fin, and $\alpha$ is the rotation angle of the nano-fin with respect to *x*-axis. The second term in Eq. 5 is cross-polarized; we refer to its normalized amplitude squared as the polarization conversion (PC) efficiency hereafter (see Methods for simulation details). The phase shift is determined by the product $(\tilde{t}_L - \tilde{t}_S)\cdot \exp(i2\alpha)$, i.e. the polarization converted light acquires a geometric phase equal to twice the rotation angle $\alpha$, whereas the group delay is *only* related to $\tilde{t}_L - \tilde{t}_S$ since $\alpha$ is frequency-independent. This provides an additional degree of freedom that allows us to decouple the required phase and group delay. In other words, in designing the individual phase shifting elements, one can first satisfy the required group delay and then adjust the rotation angle $\alpha$ to meet the required phase of every location on the metalens. Figure 2(c) shows a plot of phase as a function of frequency for a nano-fin with $l$ = 250 nm and $w$ = 80 nm with different rotation angles. The slope is with good approximation linear within a given bandwidth, and is independent of the rotation angle of the nano-fin. This property allows us to design achromatic metalenses with a large bandwidth.

To gain physical insight into the dispersion design, it is helpful to regard $TiO_2$ nano-fins as truncated waveguides. Neglecting end reflections, the phase of the transmitted light after passing through the structure at a given coordinate *r* is then given:

$$\phi(r,\omega) = \frac{\omega}{c} n_{eff} h \qquad (6)$$

where $n_{eff}$ and $h$ represent the effective index and the height of the nano-fin, respectively. Figure 2(d) shows a comparison of PC efficiency using the eigenmode solver and finite-difference time-domain (FDTD) methods (Methods). They are in good qualitative agreement, which verifies the validity of treating the nano-fins as short waveguides. The larger deviation at higher frequencies results from the excitation of higher order modes and resonances [30, 31]. The derivative of Eq. 6 with respect to angular frequency:

$$\frac{\partial\phi(r,\omega)}{\partial\omega} = \frac{h}{c} n_{eff} + \frac{h\cdot\omega}{c}\frac{\partial n_{eff}}{\partial\omega} \qquad (7)$$

yields the group delay: this is the ratio of the nanofin height to group velocity, which can be controlled by the nanofin dimensions and/or material used. Figure 2(e) shows the phases and PC efficiencies of five different elements. Their group delays were obtained using linear fitting of the curves within a bandwidth of 120 nm centered at 530 nm (see Methods for details). This ensures that the group delay of an element fulfills the requirement given by Eq. 3 and its GDD is close to zero for at least the 120 nm bandwidth being considered. However, as seen later in the text, our simulations and experimental results show that the metalens focal length is weakly dependent on wavelength beyond this bandwidth up to 670 nm. It should be noted that the achromatic metalenses implemented by this method can be highly efficient depending on the range of group delays (See Fig. S3 for a plot of PC efficiencies versus group delays for all elements available). In addition, most of the highly PC efficient nano-fins have group delays between 2 to 5 femto-seconds. This is a result of the waveguiding effect being dominant. The observed group delay values can be obtained by substituting the appropriate $n_{eff}$ between 1 and 2.4 (the refractive index of air and $TiO_2$) into the first term of Eq. 7.

**Achromatic focusing and imaging**

To demonstrate the versatility of our approach, we designed and fabricated an achromatic metalens ($n$ = 0) together with two other metalenses with $n$ = 1 and 2. They all have a NA = 0.2 at wavelength $\lambda$ = 530 nm. For $n$ = 1, i.e. a regular diffractive metalens, the phase profile was imparted by identical nanofins using the geometric phase [32]. The achromatic metalenses were designed by digitizing the required phase and group delay. Subsequently, the digitized group delays were implemented by selecting elements from a library of various nano-fin parameters. For $n$ = 2, we selected elements with dispersions from the library close to the required ones (see Methods for details). Videos showing the imparted phases for the $n$ = 0 and $n$ = 2 metalenses are provided in the Supplementary Movie 1 and 2. The measured normalized focal length shifts and the theoretically predicted values from $\lambda$ = 470 nm to 670 nm are shown in Fig. 3(a). The focal lengths were calculated by propagating the wavefronts from the meta-lens surface using Fresnel−Kirchhoff integration. Experimentally, the focal lengths at different wavelengths were obtained by measuring their intensity profiles (point spread functions) along the propagation direction ($z$-axis) of the incident beam in steps of 1 μm, as shown in Fig. 3(b) to 3(d). The $z$-coordinate corresponding to the peak intensity value gives the focal length for a given wavelength.

We also characterized the performance of these diffractive and achromatic meta-lenses in terms of their focal spot profiles (Fig. 3(e) and 3(f)). They were measured at the focal plane corresponding to an illumination wavelength of 470 nm, along the white dashed lines in Fig. 3(b) and 3(c). The diffractive metalens shows significant defocusing when the wavelength of incidence is larger than 550 nm (Fig. 3(e)). In contrast, the focal spots of achromatic metalens at different wavelengths are diffraction-limited: their Strehl ratios are larger than 0.8 and the deviations of FWHM are within 5% of the theoretical values (see Fig. S4). A real-time video showing the focal spot profiles as the incident wavelength is swept across the visible can be found in Supplementary Movie 3. It is important to note that although we only designed this achromatic metalens for a bandwidth of 120 nm centered at 530 nm, due to the negligible focal length shift compared to its depth of focus ($\frac{\lambda}{NA^2}$), the metalens maintains its focal spot profile for the entire visible spectrum from 470 nm to 670 nm. This achromatic metalens can also focus white incoherent light source under a broadband halogen lamp illumination (Fig. S5).

We also fabricated achromatic and diffractive metalenses with larger diameters (NA = 0.02, diameter = 220 μm) and compared their imaging quality. The respective focal spot profiles of the achromatic metalens are shown in Fig. 4(a). A real-time video of focal spot profile measurement is provided in Supplementary Movie 4. Figure 4(b) shows the images of a standard USAF resolution target obtained from the achromatic metalens, under various laser illumination wavelengths with a bandwidth of 40 nm (see Methods for experimental setup). This corresponds to the full-width at half-maximum (FWHM) of a typical LED light source. See Fig. S6 for similar measurements using the diffractive metalens. The test target was fixed at the focal plane corresponding to illumination at $\lambda = 470$ nm. In Fig. 4(b), a slight decrease of contrast in the images at red wavelengths is observed since the feature size of the target (~ 15 μm) is close to the diffraction limit of the achromatic metalens; there is also a decrease in efficiency of the metalens at red wavelengths. Additionally, we demonstrate white light imaging using a broadband illumination source (white light laser) from 470-670 nm. The images of all patterns from the USAF target and Siemens star are shown in Fig. 4(c) and 4(d). These images show that chromatic aberration is well-corrected even under white light illumination, and that the metalens is able to achieve high imaging quality over a few square millimeters, corresponding to a 30 degrees field of view. Note that the patterns at the center of the USAF target and the Siemens star have feature sizes smaller than the resolution of the achromatic metalens. We measured the focal spots of the metalens for different angles of incidence up to 15 degrees from 470 to 670 nm (Fig. S7). The color of these white light images has a blueish tinge because of the wavelength-dependent efficiency of the achromatic meta-lens, which peaks at 20 % around 500 nm, against a theoretically predicted value of 50% (Fig. S8(a)). The efficiency is defined as the power measured at the focal spot divided by that of circularly polarized incident light passing through an aperture with the same diameter of the metalens. This deviation likely results from fabrication errors and the coupling between metalens elements.

Our design principle can be applied to other regions of the electromagnetic spectrum [33, 34]. For example, in the infrared, one can utilize waveguide dispersion of a dielectric pillar to compensate for its material dispersion, a technique which is widely used in optical fibers and on-chip photonic circuits at telecom wavelengths. The limiting factor in realizing these achromatic metalenses with larger sizes is the range of group delays given by all possible geometric parameters and combinations of nano-fins. This can be increased by either increasing the height or by engineering resonances of the nano-fins [35-40]. Increasing the height can be realized by using multi-level lithography or two photon-polymerization fabrication techniques [41, 42]. Using single crystal silicon which can provide larger $n_{eff}$ can also increase the range of achievable group delay in the visible [43]. Stacking layers of metalenses is another feasible approach to correct not only chromatic but also monochromatic aberrations (coma, field curvature, astigmatism etc.) within a larger field of view [44-46].

**Conclusions**

By simultaneously controlling the phase and group delay, we have demonstrated dispersion-tailored metalenses in transmission over a large continuous bandwidth in the visible region. This represents a significant advance in the state-of-the-art for metalenses, which have traditionally been limited in their applications by bandwidth limitations in the visible. We have demonstrated an achromatic planar meta-lens (NA = 0.2) capable of focusing light to the diffraction-limit from 470 to 670 nm, and an achromatic meta-lens (NA = 0.02) for imaging. The planar fabrication process avoids the drawbacks associated with traditional lens-polishing techniques. Achromatic and dispersion-tailored metalenses can find numerous applications across industry and scientific research, such as in lithography, microscopy, endoscopy and virtual/augmented reality.

**Methods**

*Design and simulation.* Simulations were performed using 3D finite-difference time-domain method (FDTD and Eigenmode solutions from Lumerical Inc.). A large number of elements, each comprising of one or two nano-fins of different lengths, widths and arrangements, were simulated under illumination of circularly polarized light to build a library. Periodic and perfectly matched layer boundary conditions were used along the transverse and longitudinal directions with respect to the propagation of light. For each simulation, the far-field electric field of an element was recorded, then normalized to the field with substrate only. The polarization conversion efficiency and phase spectrum were obtained from the normalized electric field.

To design the achromatic meta-lenses, we linearly fit the phase spectrum of each element in our library at the design wavelength (530 nm) within a 120 nm bandwidth to obtain the group delay. Any element that has a R-squared value and polarization conversion efficiency less than 0.98 and 5% was dropped. The linear fitting with high R-squared value ensures that all selected elements from the library fulfill the requirement for realizing achromatic meta-surfaces with larger than 120 nm bandwidth. Our simulations and measurements of the point spread function have shown that the metalens remains achromatic with up to a bandwidth of 200 nm. Finally, for each coordinate of the achromatic metalens, the required group delay was achieved by positioning the corresponding suitably rotated element in order to satisfy the required phase profile at wavelength $\lambda$ = 530 nm. For the metalens with n = 2, we fit the phase spectrum with a quadratic polynomial to obtain group delay and GDD. Two different offset values are chosen by the particle swarm algorithm to best satisfy the required relative group delay and GDD (see Fig. S9 for details).

*Measurement.* The meta-lenses were characterized using a custom-built microscope mainly consisting of a tunable laser (SuperK and Varia from NKT) and an Olympus objective lens (100×, NA = 0.95) paired with a tube lens (focal length *f* = 180 mm) to form an image on a camera. The objective lens was moved 1 um per step by a motorized stage (Märzhäuser, MA42) controlled by Labview to record the intensity distributions along the beam propagation direction on a monochromatic camera (Thorlabs, DCC1545M) for different wavelengths. Subsequently, these images are stacked to obtain Fig. 3(b) to 3(d). For measuring the focal spot profiles shown in Fig. 3(e) and 3(f), the camera was replaced by another with smaller pixel size (Edmund, EO-5012). For all measurements, a pair of crossed circular polarizers were used to reduce background noise. A schematic of the experimental setup and details for Fig. 4 can be found in Fig. S10.

**Supplementary Materials:**

Figure S1-S10

Movies S1-S4

**Author contributions:**

W. T. C. and F.C. conceived of the study. A. Y. Z. fabricated the metalenses. W. T. C., V. S., M. K., Z. J. and E. L. performed simulations and developed codes. W. T. C., A. Y. Z. and E. L. measured the metalenses. W. T. C., A. Y. Z., M. K. and F. C. wrote the manuscript. All authors discussed the results and commented on the manuscript.

**Competing financial interests**

The authors declare no competing financial interests

**Acknowledgements**

This work was supported by the Air Force Office of Scientific Research (MURI, grant# FA9550-14-1-0389 and grant# FA9550-16-1-0156). This work was performed in part at the Center for Nanoscale Systems (CNS), a member of the National Nanotechnology Coordinated Infrastructure (NNCI), which is supported by the National Science Foundation under NSF award no. 1541959. CNS is part of Harvard University. Federico Capasso gratefully acknowledges a gift from Huawei Inc. under its HIRP FLAGSHIP program. We thank Yao-Wei Huang and Jared Sisler for their helps in measurement and in simulations, respectively.

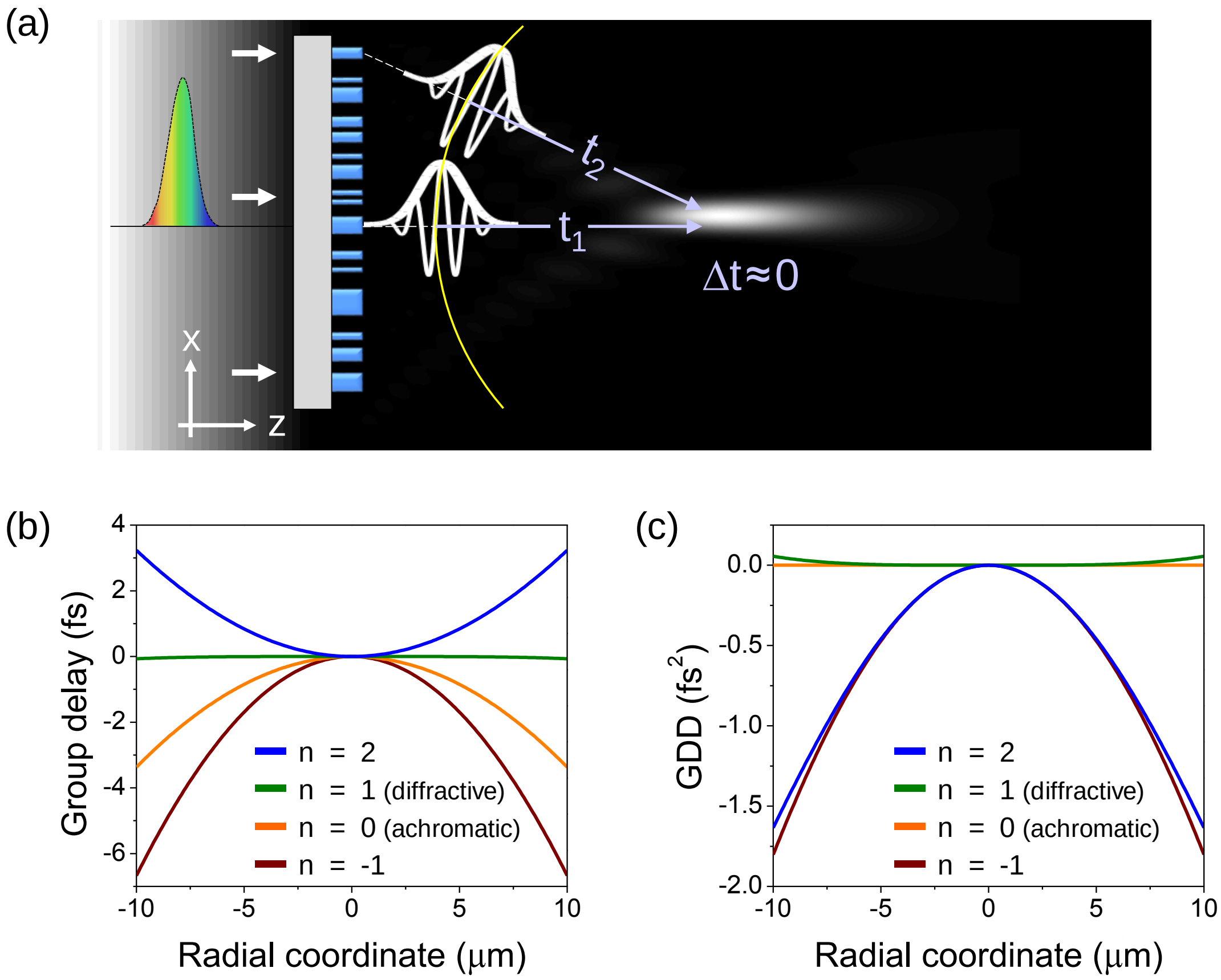

**Figure 1: Dispersion engineering of achromatic and chromatic metalenses.** (a) Schematic of an achromatic metalens. To realize achromatic focusing, the phase profile of $\varphi(r,\omega)$ must satisfy Eq. 1 in the text. The metalens is designed to provide spatially dependent group delays such that wavepackets from different locations arrive simultaneously at the focus. The yellow line shows the spherical wavefront. (b) Required relative group delays as a function of metalens coordinate. The focal length is parametrized as $F(\omega)=k\omega^n$. Depending on the value of n, the metalens can be designed as achromatic (n = 0) or chromatic with focal length inversely proportional to wavelength (n = 1,

dispersion similar to Fresnel lenses) or proportional to wavelength (n = -1). The case of n = 2 exhibits stronger dispersion. These metalenses have a diameter of 20 μm and a focal length of 63 μm at λ = 530 nm. (c) Required relative group delay dispersion of the same metalenses.

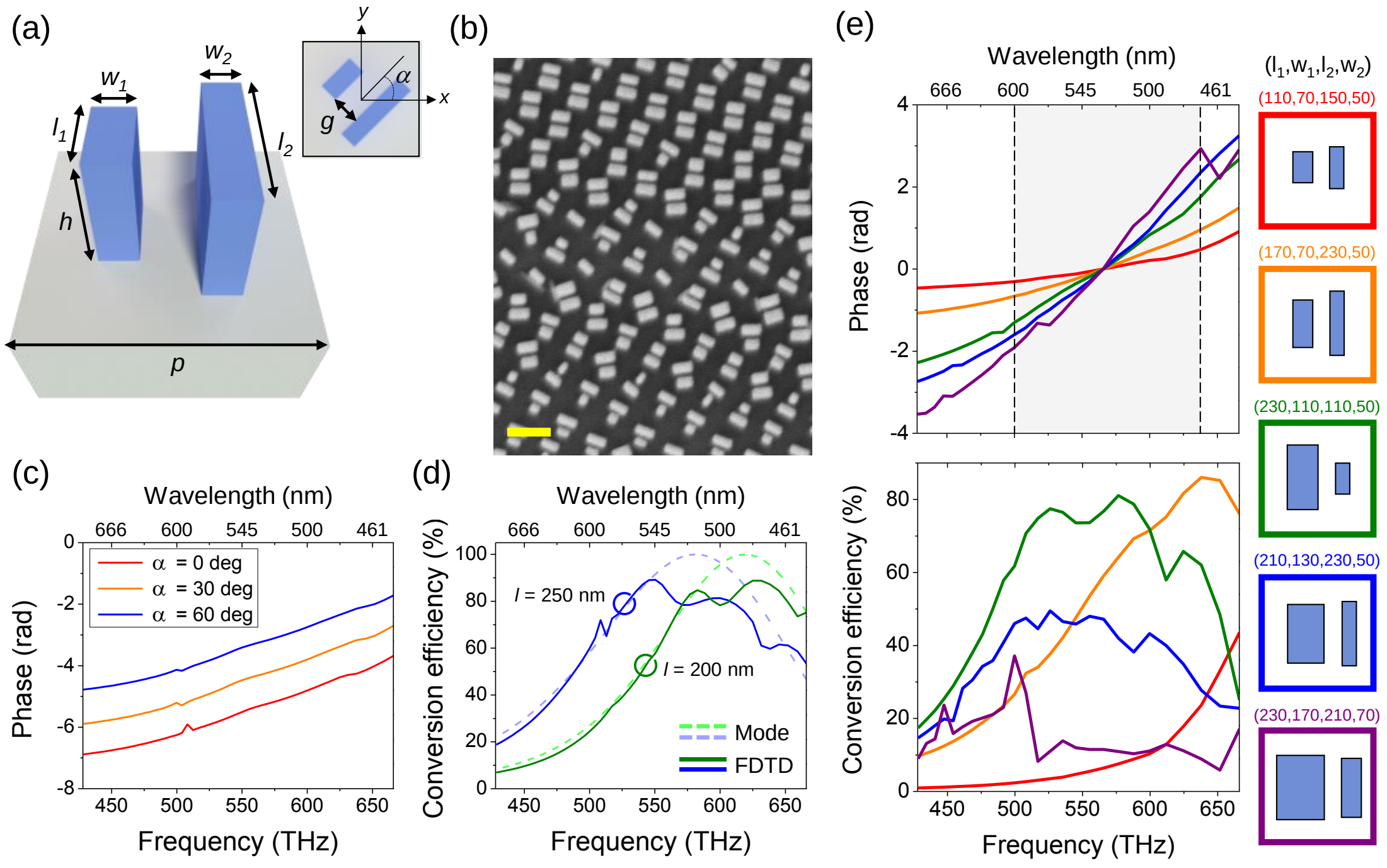

**Figure 2: Optical properties of nanofins and scanning electron micrograph.** (a) Schematic of a metalens element. The element consists of one or more $TiO_2$ nanofins of varying dimensions but equal height *h* = 600 nm, evenly spaced by a distance *p* = 400 nm. The gap between nanofins is *g* = 60 nm. The length *l*, width *w*, height *h* and rotation angle $\alpha$ are also shown; the subscripts denote the left and right nanofin, respectively. The nanofins are rotated with respect to the center of the square (400 × 400 $nm^2$). (b) Scanning electron micrograph of a region of a fabricated metalens. Scale bar: 500 nm. For more images of the metalens, see Fig. S2. (c) and (d) Simulation results for a single nanofin. (c) Phase plots as a function of frequency for different rotation angles for nanofins with *l* = 250 nm and *w* = 80 nm. (d) A comparison of polarization conversion efficiency for different

nanofin lengths from FDTD calculations (solid lines) versus Mode Solutions (dashed lines). The lengths of the nanofins are labeled; they have a constant width $w$ = 80 nm. (e) Phase spectra and polarization conversion efficiencies for five different elements showing the tunability of the group delay by changing the lengths and widths of nanofins. The shaded region marks the design bandwidth of 120 nm. Each colored curve corresponds to its element schematically shown on the right. The parameters ($l_1$, $w_1$, $l_2$, $w_2$) of each nanofins are labelled; all dimensions are in nanometers. The elements in the colored squares are located at different radial positions from the edge of the metalens (red square) to the center (purple square), such that the corresponding group delay (slop of the phase versus angular frequency plot) increases from the edge to the center. This ensures achromatic focusing as illustrated in Fig. 1(a).

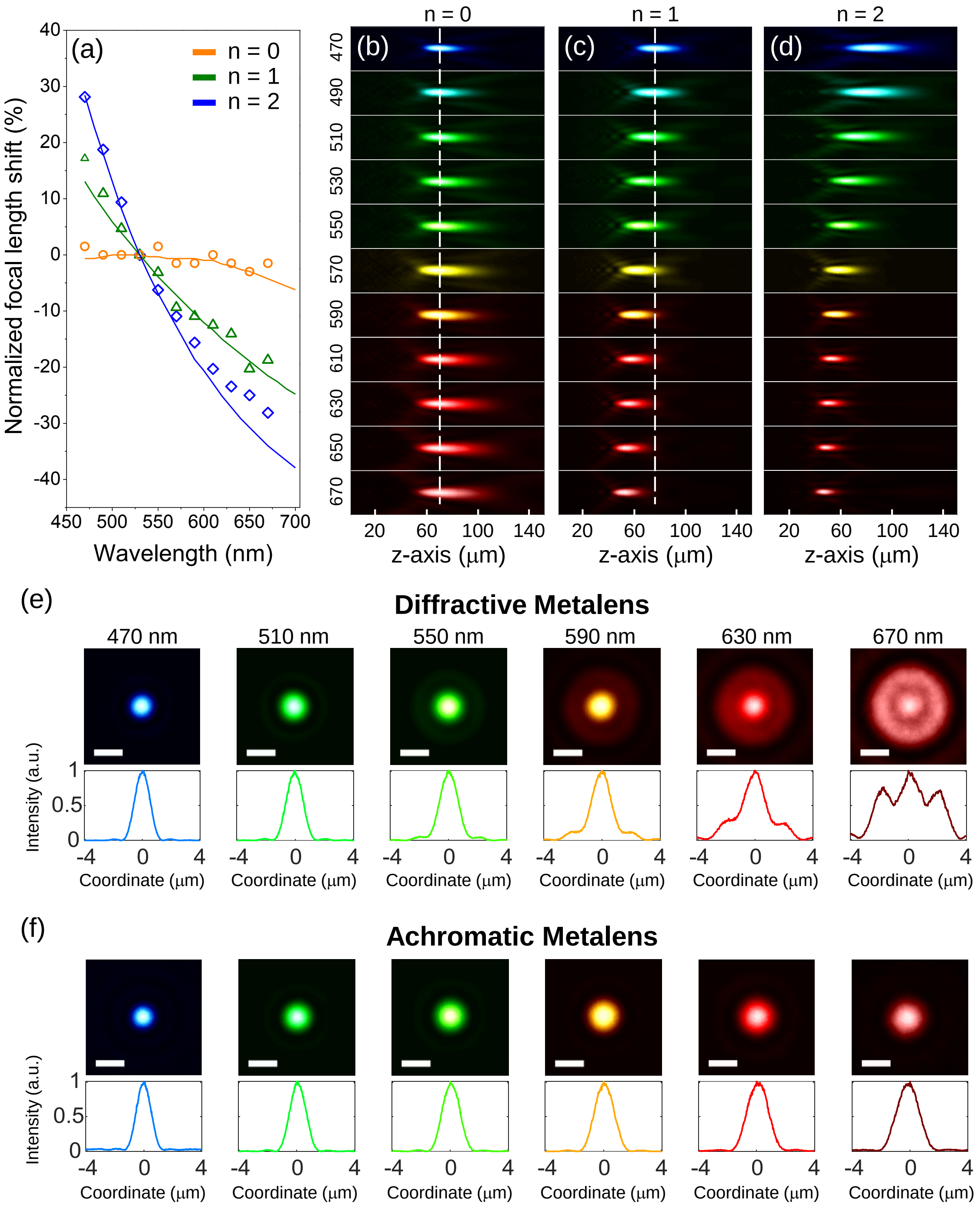

**Figure 3: Measured focal length shifts and intensity distributions of metalenses.** (a) Experimentally measured normalized focal length shifts (symbols) compared to simulations (lines) for metalenses with different values of *n*. Incident light was collimated

and at normal incidence. The metalenses were designed at wavelength $\lambda$ = 530 nm with NA = 0.2 and focal length of 63 μm. (b) to (d) Measured intensity distributions in linear scale (in false colors corresponding to their respective wavelengths) in the *x*-*z* plane. (b), (c) and (d) correspond to metalenses with $n$ = 0 (achromatic), 1 and 2, respectively. The wavelengths of incidence are denoted on the left. The direction of incidence is towards the positive *z*-axis. (e) and (f) Normalized intensity profiles along the white dashed lines of (c) and (b) for diffractive and achromatic metalenses, respectively. The white dashed lines pass through the center of focal spots in case of $\lambda$ = 470 nm. Scale bar: 2 μm.

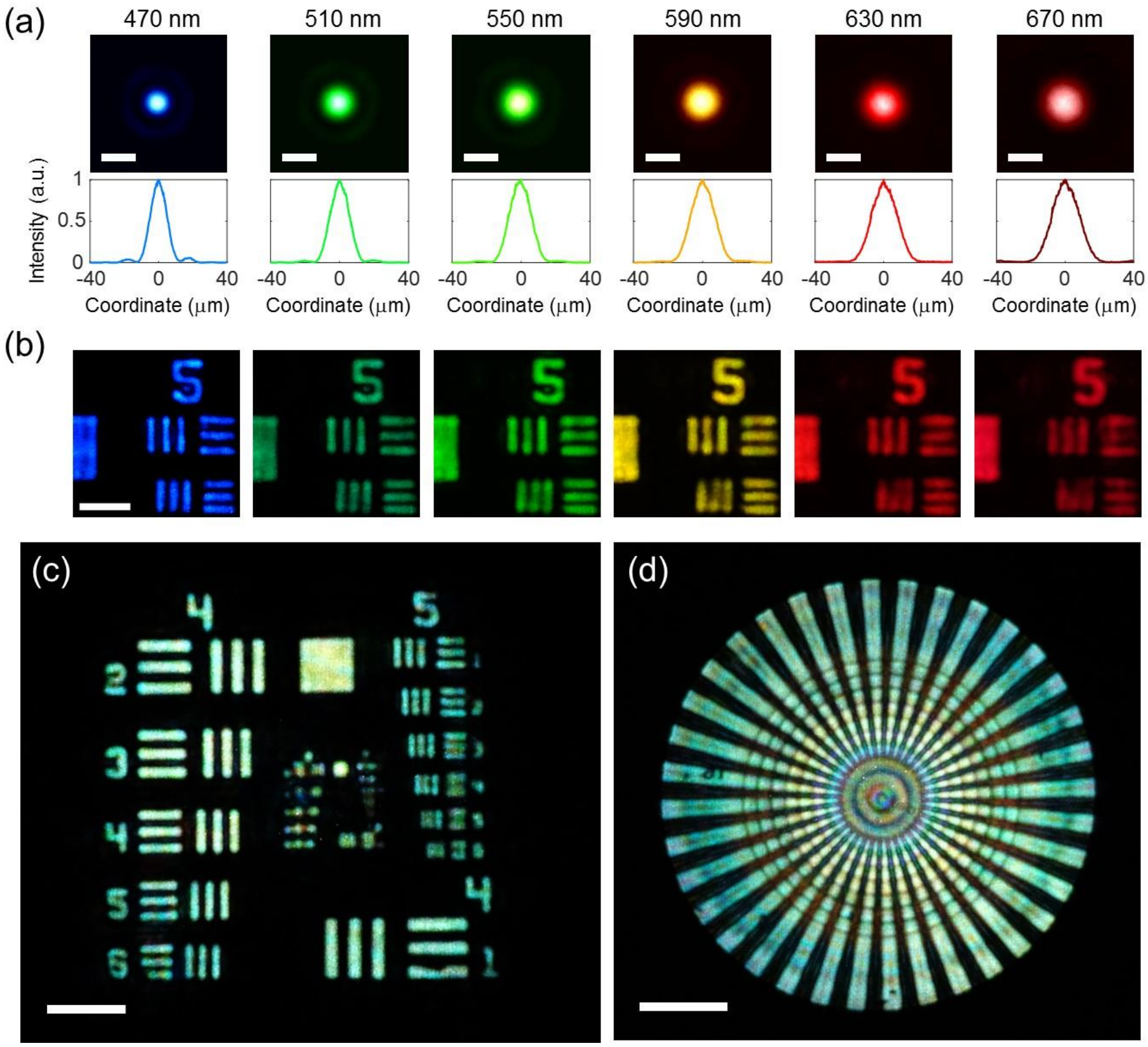

**Figure 4: Focal spot profiles and imaging using an achromatic metalens under different illumination wavelengths.** The achromatic metalens has a diameter of 220 μm and NA of 0.02. The light source is a supercontinuum laser with tunable center wavelength and bandwidth. (a) Experimentally measured focal spot profiles. The illumination wavelengths with about 5 nm bandwidth are denoted on the top. Scale bar: 20 μm. (b) Images of 1951 United States Air Force resolution target formed by the achromatic

metalens. The resolution target (Thorlabs, R1L1S1N) was fixed at the focal plane corresponding to an incident wavelength $\lambda$ = 470 nm. The line widths of upper and lower bars are 15.5 and 14 μm, respectively. Scale bar: 100 μm. (c) Images of the same Air Force target using the achromatic metalens under an illumination bandwidth of 200 nm centered at 570 nm. (d) The image of the Siemen star pattern under the same illumination condition in (c). The outer radius is 500 μm. The metalens provides about 50× magnification. Scale bar: 200 μm.